\documentclass[twocolumn,useAMS,usenatbib]{mn2e}

\usepackage{epsf}
\usepackage{bm}
\usepackage{amsfonts}
\usepackage{latexsym}
\usepackage[latin1]{inputenc}
\usepackage{amsmath}
\usepackage{epsfig}
\usepackage{amsbsy}

\def\X{{\mathrm{x}}}
\def\Y{{\mathrm{y}}}

\def\b{{\mathrm{B}}}

\def\n{{\rm n}}
\def\p{{\rm p}}

\def\v{{\rm v}}

\def\wid{w_i^{\Y\X}}
\def\wju{w^j_{\Y\X}}
\def\wjd{w_j^{\Y\X}}

\def\N{\mathrm{c}}
\def\({\left(}
\def\){\right)}
\def\[{\left[}
\def\]{\right]}
\def\eps{\varepsilon}

\def\mut{\widetilde{\mu}}
\newcommand{\ncd}{\newcommand}
\ncd{\dell}{\partial}
\ncd{\nfrac}[2]{\left(\frac{n_{#1}}{n_{#2}}\right)^2}
\ncd{\pc}{\check{p}}
\ncd{\rhoc}{\check\rho}
\ncd{\betac}{\check\beta}
\ncd{\muc}{\check\mu}
\ncd{\Oc}{\check\Omega}
\ncd{\ec}{\check\epsilon}
\ncd{\tsfrac}[2]{{\textstyle\frac{#1}{#2}}}

\newcommand{\be}{\begin{equation}}
\newcommand{\ee}{\end{equation}}
\newcommand{\beq}{\begin{equation}}
\newcommand{\eeq}{\end{equation}}
\newcommand{\bear}{\begin{eqnarray}}
\newcommand{\eear}{\end{eqnarray}}

\ncd{\bldeta}{\boldsymbol{\eta}}
\ncd{\bldone}{\mathbf{1}}
\ncd{\blds}{\mathbf{s}}
\ncd{\bldk}{\mathbf{k}}
\ncd{\blde}{\mathbf{e}}

\ncd{\abs}[1] {|#1|}
\ncd{\ubold}{\mathbf u}
\ncd{\Abold}{\mathbf A}
\ncd{\Bbold}{\mathbf B}
\ncd{\Mbold}{\mathbf M}
\ncd{\lagom}{\hspace{.6pt}}
\ncd{\muk}{k}
\ncd{\lagomdot}{{\mbox{\large$\cdot$}}}

\ncd{\stil}{\tilde{s}}
\ncd{\ftil}{\tilde{f}}
\ncd{\Otil}{\tilde{\Omega}}

\ncd{\D}{\mathcal{D}}
\ncd{\Qcal}{\mathcal{Q}}
\ncd{\Jcal}{\mathcal{J}}
\ncd{\Ecal}{\mathcal{E}}
\ncd{\Wcal}{\mathcal{W}}
\ncd{\Xcal}{\mathcal{X}}
\ncd{\Ycal}{\mathcal{Y}}
\ncd{\Bcal}{\mathcal{B}}
\ncd{\Acal}{\mathcal{A}}
\ncd{\Scal}{\mathcal{S}}
\ncd{\Ccal}{\mathcal{C}}

\ncd{\vt}{v_{t\perp t}^{\,2}}

\ncd{\psimap}{\Psi}
\ncd{\psivar}{\psi}

\ncd{\Ap}{(r\psivar)'}
\ncd{\Adot}{(r\psivar)^{\lagomdot}}
\ncd{\Bp}{(r^{-1}\varphi)'}
\ncd{\Bdot}{(r^{-1}\varphi)^{\lagomdot}}

\ncd{\ela}{\left(1-\frac{2m}{r}\right)}

\ncd{\shm}{S}
\ncd{\shmtwoD}{\mathcal{\shm}}
\ncd{\lie}{\mathcal{L}}
\ncd{\brk}{\mathrm{max}}
\ncd{\fgauge}{f_\mathrm{G}}
\ncd{\I}{\mathrm{c}}
\ncd{\f}{\mathrm{f}}

\title[Lagrangian perturbations of the neutron star crust]{Lagrangian perturbation theory for a superfluid immersed in an elastic neutron star crust}

\author[N. Andersson, B. Haskell \& L. Samuelsson]{N. Andersson$^1$, B. Haskell$^{1,2}$ \& L. Samuelsson$^{3,4}$ \\ \\
$^1$ School of Mathematics,
University of Southampton, Southampton SO17 1BJ, United Kingdom \\
$^2$Astronomical Institute ``Anton Pannekoek'', University of Amsterdam, Science Park 904, 1098 XH Amsterdam, Netherlands\\
$^3$ Department of Physics, Ume\aa\ University, SE-901 87 Ume\aa, Sweden \\
$^4$ Nordita, Roslagstullsbacken 23, SE-106 91 Stockholm, Sweden}

\begin{document}

\maketitle

\begin{abstract}
The inner crust of mature neutron stars, where an elastic lattice of
neutron-rich nuclei coexists with a neutron superfluid, impacts on a
range of astrophysical phenomena. The presence of the superfluid is
key to our understanding of pulsar glitches, and is expected to
affect the thermal conductivity and hence the evolution of the
surface temperature. The coupling between crust and superfluid must
also be accounted for in studies of neutron star dynamics,
discussions of global oscillations and associated instabilities. In
this paper we develop Lagrangian perturbation theory for this
problem, paying attention to key issues like superfluid entrainment,
potential vortex pinning, dissipative mutual friction and the star's
magnetic field. We also discuss the nature of the core-crust
interface. The results provide a theoretical foundation for a range of
interesting astrophysical applications.
\end{abstract}

\section{Motivation}

The crust of a mature neutron star shields the high-density fluid
core from the thin atmosphere. It provides a heat-blanket that
determines the link between the neutrino-driven cooling of the core
and the evolution of the observed surface temperature
\citep{heatblanket}. The physics of the crust is also of key
importance for neutron star dynamics. The presence of a superfluid
component, and the associated rotational vortices, is central to our
understanding of pulsar glitches \citep{espinoza}. The coupling
between the neutron-rich nuclei in the crust and the superfluid
neutrons also affect global oscillations involving the crust, as in
the case of the quasi-periodic oscillations observed in the tails of
magnetar flares \citep{magnetars}. On the one hand, our
understanding of the crust physics is quite good. The equation of
state for matter has been modelled in detail. In particular, we have
a clear idea of how the composition changes with density which
allows us to work out the elastic properties of the lattice of
nuclei \citep{chamelLR}. This also provides us with the fraction of
neutrons that have dripped out of nuclei and which may considered
free to move relative to the lattice. However, our understanding of
the dynamics of the coupled crust-superfluid system is quite poor.
This is somewhat surprising given the relevance of the involved
issues (e.g. in the context of glitches), but it is nevertheless the
case.

To make progress we need to develop a theoretical framework that
allows us to model the dynamics of the inner-crust region. Such a
model has to consider the key physics, like elasticity, the magnetic
field and superfluidity. It must account for the presence of
superfluid vortices, potential pinning and mutual friction. It is
also important that the model is adaptable, in order to faciliate
further development concerning, for example, finite temperature
effects and heat propagation. This is obviously a major challenge.
The aim of the present work is to develop  a ``complete'' model for
linear perturbations of a (moderately) realistic neutron star crust.
Working within the Lagrangian perturbation framework (the natural
way to view the perturbed crust, and a necessity if one wants to
consider the stability of the system from the formal point of view
\citep{fs78}), we provide a two-component model that accounts for
the key physics. The model also complements previous work on the
outer-core region \citep{kirsty} by discussing the role of the
superfluid entrainment and the magnetic field \citep{maglag}. We
also consider the conditions that need to be implemented at the
crust-core transition. The theory is developed to the point where
the results can be applied to a range of interesting astrophysics
problems. Having said that, there is massive room for future
refinements as our understanding of the relevant physics improves.

\section{The two-fluid model}

The conditions in the inner crust, with an elastic lattice of
increasingly neutron-rich nuclei immersed in a neutron superfluid
has (at least) two dynamical degrees of freedom; the superfluid
neutrons may flow relative to the lattice. In essence, this is an
example of a two-``fluid'' system, although in this case one of the
components is also affected by elastic restoring forces. Our model
for this system is based on the variational approach to multi-fluid
dynamics developed by \citet{prix} and \citet{monster} (see also
\citet{brynform}, which represents the Newtonian limit of the
fully relativistic convective variational model designed by \citet{carter}, see also
\citet{livrev}. The elastic sector builds on the relativistic model
developed by \citet{maxlars}, and the inclusion of superfluidity
follows the strategy set out by \citet{casa,cacho}.

\subsection{Hydrodynamics, magnetic fields and elasticity}

In order to model the conditions in the inner crust of a neutron
star, we take as the starting point the equations for two-fluid
hydrodynamics \citep{prix,monster,brynform}. Assuming that the
individual components are conserved (and working in a coordinate
basis where vectors are represented by their components, with
indices $i,j,k$ as usual), we then have the usual conservation laws
for the number densities $n_\X$, where $\X$ is a constituent index
labeling the components,
\be
\partial_t n_\X + \nabla_i (n_\X v_\X^i) = 0 \ .
\label{conti}\ee
Note that a repeated species index x does not imply summation, in contrast to the spatial indices like $i$
for which the Einstein summation convention applies.
 In the outer core of a neutron star, the
distinction between the two components is fairly clear. On the one
hand, we have the superfluid neutrons. On the other hand, we have a
conglomerate of (most likely superconducting) protons and electrons.
On scales larger than the electron screening length and timescales
longer than the inverse of the plasma frequency \citep{mendell,supercon}, these form a single, charge neutral, fluid. The two-fluid
model simply distinguishes the neutrons (represented by $n_\n$, say)
from the protons/electrons (given by $n_\p$). The situation in the
crust is similar yet different. At densities beyond neutron drip, some neutrons
remain bound in nuclei but there is also a ``gas'' of free neutrons.
The assignation of neutrons to each component follows from the
equation of state, once the nature of the ions in the lattice is
established \citep{chamelLR}. However, if we consider the dynamics of the system it
is not clear to what extent the ``confined'' neutrons are able to
move \citep{cacha}. This depends on how strongly bound they are, to what extent
they can tunnel through the relevant interaction potentials
etcetera. The upshot is that one can choose to work in different
chemical ``gauges''. This issue has been discussed in detail by
\citet{cacha} (and we will return to it later). The choice of chemical gauge affects the
interpretation of the involved quantities (number densities, etc),
but  the two-fluid model remains unchanged conceptually. To make a
distinction from the outer core problem, we will refer to the two
components in the crust as ``free'' neutrons, with density $n_\f$,
and ``confined'' baryons, represented by $n_\I$. This notation may
represent a slight bias towards to description advocated by \citet{cacha} and \citet{casa}, but at this point we
basically want to keep the options open by not linking the
discussion too much to established results for the outer core.

The ``free'' neutrons can flow relative to the crust lattice once the system cools below the transition to
$^1S_0$ neutron superfluidity (see \citet{nphysa} for typical transition temperatures). We then have two coupled
 equations of momentum balance. Assuming that the large scale system comprises a sufficient number of
 quantized vortices that macroscopic averaging is meaningful (this should, indeed,
 be the case for all astrophysical systems of interest) the two momentum equations can be written \citep{prix}
\be
(\partial_t + v_\X^j \nabla_j ) (v^\X_i+\varepsilon_\X w^{\Y\X}_i) +\nabla_i (\tilde{\mu}_\X+\Phi)\\ 
+ \varepsilon_\X w^j_{\Y\X} \nabla_i v^\X_j 
= f^\X_i/\rho_\X
\label{Eulers}\ee
where  $\X=\{\f,\I\}$, the velocities are $v_\X^i$, the relative velocity is defined as $w_{\X\Y}^i = v_\X^i-v_\Y^i$ and
 $\tilde{\mu}_\X=\mu_\X/m_\X$ represents the
chemical potential scaled to the nucleon mass (we will assume that the neutron and proton masses are equal, $m_\N=m_\n=m_\b$). The mass densities are
given by
$\rho_\X=m_\b n_\X$ and $\Phi$ represents the gravitational potential, which means that we have
\be
\nabla^2 \Phi =  4 \pi G (\rho_\f + \rho_\I)
\ee
and, finally, the parameter
$\varepsilon_\X$ encodes the entrainment. The entrainment can be expressed in terms of a single parameter
$\alpha$ such that \citep{prix}
\be
\rho_\X \varepsilon_\X = 2 \alpha
\ee
To close the system, we need to provide an equation of state. In the present formalism, the equation of state takes the
form of an energy functional $E$, the functional form of which determines the chemical potentials
\be
\mu_\X = \left( {\partial E \over \partial n_\X}\right)_{n_\Y,w^2} \ ,
\ee
and the entrainment parameter
\be
\alpha = \left( {\partial E \over \partial w^2}\right)_{n_\X,n_\Y} \ .
\ee

The forces on the right-hand side
of (\ref{Eulers}) can be used to represent various other interactions, including dissipative terms \citep{monster,brynform}. If we focus on the
conditions in the neutron star crust, then we need to account for elasticity and the large-scale magnetic field.
The latter is described by the usual electromagnetic Lorentz
force. That is, we have
\be
f_i^\I  = \frac{1}{c} \epsilon_{ijk} J^j B^k
\label{lorentz0}
\ee
Eliminating the total current with the help of Amp\'ere's law,
i.e.  $J^i = (c/4\pi) \epsilon^{ijk} \nabla_j B_k$,
this becomes
\be
f_i^\I = { B^j \over 4\pi } (\nabla_j B_i - \nabla_i B_j )\
\label{lorentz1}
\ee
In order to use this in \eqref{Eulers} we also need to know $n_\I$, i.e., to what extent the baryons are affected
by the magnetic field. It is natural to assume that all baryons that are confined to nuclei are involved, but
the exact meaning of this is (as we will discuss later) somewhat fuzzy in a dynamical situation in regions of the stars
where some of the neutrons are free.

The elastic force is different in that it only serves to restore deviations from a relaxed state of the lattice.
This means that it is natural to discuss elasticity at the linear perturbation level. We will do this later, in Section~4.2.
In that discussion, we will assume that the background configuration is relaxed, in which case there is
no leading order elastic force in \eqref{Eulers}. This assumption is not quite realistic; the crust of an astrophysical
neutron star is likely to be strained due to the regular electromagnetic spin-down of the system.
It is, in principle, straightforward to account for this strain but in the interests of clarity we have chosen
not to do so here.

Before we move on, it is worth noting that the incorporation of elasticity requires us to track given
``fluid'' elements relative to the relaxed configuration. This motivates us to
use a Lagrangian framework. This approach is, of course, advantageous for a number of reasons.
In particular, if we are interested in considered rotating neutron stars. Our model for the inner crust
builds on the two-fluid perturbation framework developed by \citet{kirsty}, adds the magnetic field according to the
analysis of \citet{maglag} and provides a model for elasticity which represents the Newtonian limit of
the theory developed by \citet{casa}.

\subsection{Superfluidity and vortex dynamics}

Let us turn to our attention to the superfluid aspects of the problem. Doing this, we note that
\eqref{Eulers} accounts for the presence of a (macroscopically averaged) vortex array. In order to
discuss issues concerning, for example, vortex pinning and mutual friction it is useful to
consider \eqref{Eulers} in more detail. Following \citet{supercon} we introduce the momentum
\be
p^\f_i = m \left( v^\f_i + \varepsilon_\f w^{\I\f}_i \right)
\ee
In a superfluid, the momentum arises as the gradient of the condensate wavefunction. The upshot of this is that the superfluid is irrotational. However, this is only true on the microscopic scale. On the scale of hydrodynamics, the superfluid can rotate by forming vortices and when these are averaged over the system mimics a system with bulk rotation (as evidenced by \eqref{Eulers}). The rotation is, however, quantized and  the vorticity is given by
\be
\mathcal{W}_\f^i = { 1 \over m} \epsilon^{ijk} \nabla_j p^\f_k = n_\v \kappa_\f^i
\label{circolo}\ee
where  $n_\v$ is the number of vortices per unit surface area and $\kappa_\f^i = \kappa \hat{\kappa}^i$ (with $\hat{\kappa}^i$ a unit vector along the direction of the vortex array and $\kappa=h/2 m \approx 2 \times 10^{-3}$ cm$^2$ s$^{-1}$ the quantum of circulation).

From equation (\ref{circolo}) one can derive the equation that governs the vorticity.
Assuming that the vortex density is conserved we have
\be
\partial_t n_\v +\nabla_i (n_\v v^i_\v) =0\label{circolo2}
\ee
where $v^i_\v$, in fact, \underline{defines} the macroscopically averaged vortex velocity. The fact that the vortices move with
 $v^i_\v$ also means that (in terms of the Lie derivative $\mathcal{L}_{v_\v}$ along $v_\v^i$) we have
\be
\left( \partial_t + \mathcal{L}_{v_\v} \right) \kappa^i = 0  \label{vortlie}
\ee
Given these relations, it follows that
\be
\partial_t \mathcal{W}^\f_i+\epsilon_{ijk} \nabla^j \left( \epsilon^{klm}  \mathcal{W}^\f_l v^\v_m \right) =0\label{circolo1}
\ee

In order to make contact with the macroscopic description, we now rewrite the relevant Euler equation, c.f. \eqref{Eulers}, as
\be
\partial_t p^\f_i + \nabla_i \left( \mu_\f - {m\over 2} v_\f^2 + v_\f^j p^\f_j \right) - m \epsilon_{ijk}
v_\f^j \mathcal{W}_\f^k =  f^\f_i/n_\f
\label{eulern}\ee
This expression makes it clear that, in the absence of vortices and external forces, the superfluid motion follows from the gradient of a scalar potential.
Moreover, it is now easy to compare \eqref{circolo1} and \eqref{circolo2}. As discussed by \citet{supercon}, we then find that
the models are consistent provided we account for the ``Magnus force'' on the right hand side of equation \eqref{eulern}.
This force takes the form
\be
\frac{f_i^\f}{\rho_\f}= n_\v \epsilon_{ijk} \kappa^j (v_\f^k-v_\v^k)
\label{magnus}\ee
and an equal and opposite force will act on the vortex array. In the simple case of a single condensate at zero temperature, force balance on the vortices
requires them to flow with $v_\f^i$ (here and in the following we ignore the inertia of the vortices \citep{mendell}). In a more general situation,
 we can still use the above strategy to account  the forces that act on the vortices. We ``simply'' solve the force balance equation for the vortices for
$v_\v^i$ and use the result in \eqref{magnus}. In the case of resistive scattering off of the vortex cores, e.g. by phonons, this leads to the
usual representation of the vortex-mediated mutual friction \citep{trev}. We will now consider this problem in the context of the crust.

\section{Mutual friction and vortex pinning}

To discuss the various vortex forces, we takes as our starting point the equation of force balance for a
single vortex;
\be
\epsilon^{ijk} \hat{\kappa}_j (v^\v_k - v^\f_k) + \mathcal{R} (v_\I^i - v_\v^i) + \mathcal{F}^i = 0
\ee
This accounts for (i) the Magnus force, (ii) a resistive friction
associated with the normal component (e.g. nuclei, electrons and phonons in the crust), with coefficient $\mathcal{R}$, and (iii) a general force which we leave unspecified
at this point. This force will later be taken to represent the ``pinning'' of vortices to the nuclei in the crust.
Note that the different terms  all have the dimension of velocity. In order to obtain
an expression for the force per unit length of a vortex, as required in the macroscopic Euler equations, we need to multiply by $\rho_\f \kappa n_\v$.
This step assumes that the vortices form a recti-linear array (the usual Abrikosov lattice). It is the simplest set-up, essentially since it make the averaging over vortices trivial,
but there is no guarantee that this situation prevails in a real system. In particular, it may be relevant to worry about the
formation of vortex tangles and superfluid turbulence \citep{trevturb,peralta}. We will not consider this problem here. Neither will we account for contributions like the
vortex tension or the elasticity of the vortex array. 
These effects are readily incorporated in our framework (see \citet{haskell} for a recent discussion of the elasticity of the vortex lattice) but we leave them out in the interest of clarity.
Finally, we assume that there are no force contributions along $\hat{\kappa}^i$, the direction of the vortex axis.

Solving for the vortex velocity in the standard way, we find
\be
v_\v^i = v_\N^i + { 1 \over 1 + \mathcal{R}^2} \left(
\mathcal{R} f^i + \epsilon^{ijk} \hat{\kappa}_j f_k \right)
\label{vv}\ee
where
\be
f^i = \epsilon^{ijk} \hat{\kappa}_j w^{\N \f}_k + \mathcal{F}^i
\ee

Now, the reaction force acting on the normal component will be
\begin{multline}
{ f_\N^i \over \rho_\f \kappa}  = -  \mathcal{R} (v_\N^i - v_\v^i)  - \mathcal{F}^i
\\
= { \mathcal{R} \over 1 + \mathcal{R}^2} \left(
\mathcal{R} f^i + \epsilon^{ijk} \hat{\kappa}_j f_k \right)
+ { 1 \over
\left(1 + \mathcal{R}^2 \right) } \hat{\kappa}^i \left( \hat{\kappa}_j \mathcal{A}^j \right) - \mathcal{F}^i \\
= { \mathcal{R}\over 1 + \mathcal{R}^2 } \epsilon^{ijk} \hat{\kappa}_j \left[ \mathcal{R} w^{\N\n}_k + \mathcal{F}_k \right] - {1 \over 1 + \mathcal{R}^2 }\perp^i_j
\left[ \mathcal{R} w_{\N\f}^j+ \mathcal{F}^j \right]
\end{multline}
where we have defined the projection orthogonal to the vortices;
\be
\perp^i_j = \delta^i_j - \hat{\kappa}^i \hat{\kappa}_j
\ee
The force acting on the neutrons will naturally be equal and opposite;
\be
{ f_\f^i \over \rho_\f \kappa}=-\epsilon^{ijk} \hat{\kappa}_j (v^\v_k - v^\f_k) =-{ f_\N^i \over \rho_\f \kappa}
\ee
For later convenience, it is  natural to introduce basis vectors along the macroscopic relative flow (in the plane orthogonal to the vortex).
That is, we use
\be
\hat{w}^i = w^i/w 
\ee
where
\be
w^i = \perp^i_j w_{\N\f}^j  \quad \mbox{ and } \quad w^2 = \left( \perp^i_j  w_{\N\f}^j \right) w^{\N\f}_i
\ee
together with the decomposition
\be
\mathcal{F}^i = a_\parallel \hat{w}^i + a_\perp \epsilon^{ijk} \hat{\kappa}_j \hat{w}_k
\label{acomp}\ee
This means that we get
\begin{multline}
{ \left( 1 + \mathcal{R}^2 \right) \over \rho_\n \kappa} f_\N^i \\
=   \left[ \mathcal{R}^2 w + a_\parallel \mathcal{R} - a_\perp \right] \epsilon^{ijk} \hat{\kappa}_j \hat{w}_k \\
-  \left[ \mathcal{R} w + a_\perp \mathcal{R} + a_\parallel \right]  \hat{w}^i
\end{multline}
Expressing the result in the usual form \citep{trev} we have
\be
{ f_\N^i \over \rho_\f \kappa}
= w \mathcal{B}'_\mathrm{eff} \epsilon^{ijk} \hat{\kappa}_j \hat{w}_k - w \mathcal{B}_\mathrm{eff} \hat{w}^i
\label{f1}\ee
where
\be
\mathcal{B}_\mathrm{eff} = { 1 \over 1 + \mathcal{R}^2 } \left[ \mathcal{R} + { a_\parallel + a_\perp \mathcal{R} \over w } \right]
\label{bb1}\ee
and
\be
\mathcal{B}'_\mathrm{eff} = { 1 \over 1 + \mathcal{R}^2 } \left[ \mathcal{R}^2 + { a_\parallel \mathcal{R} - a_\perp \over w } \right]
\label{bb2}\ee
The first term in each bracket represents the standard mutual friction. The second terms illustrate how a pinning force may be accounted for
in the macroscopic multi-fluids model. It is worth noting that this model can also be applied to the problem of (potentially strong) interaction
between neutron vortices and proton fluxtubes in the outer core of a neutron star, c.f. \citet{supercon}.

\subsection{Perfect pinning}

Having discussed the general model, we are equipped to consider the limiting case of perfect ``pinning''. The interaction between the
vortex lines and the crustal nuclei may be strong enough to ``pin'' the vortices and force them to move with along with the crust \citep{pierre}.
This has profound
implications for the macroscopic dynamics of the system. Given that the vortex lines are no longer free to move, the superfluid neutrons cannot spin down (or up).
Hence, a lag will build up between the two components as the crust slows down due to magnetic braking. When this lag develops, the Magnus force will tend to
push the vortices out (or in),
c.f. \eqref{magnus}. Eventually, the force will be strong enough  to overcome the pinning and break the vortices free. This leads to a transfer of angular momentum,
that could explain
large pulsar glitches (see \citet{andrea} for a recent discussion). Vortex pinning may also have a severe effects on neutron star precession. By acting as a gyroscope, the pinned vortices are expected to lead to
extremely short period precession, of the order of the rotation period (rather than the several months to years period expected from a typical crustal deformation)
\citep{jones,linkprec}.  While this general picture is
supported by a range of theoretical models, we are still quite far from a detailed understanding of the nature and strength of vortex pinning. However,
for the present study it is sufficient to assume that a pinning force is acting.

Let us assume that there is ``perfect'' pinning, $v_\v^i=v_\N^i$. In this  case the equation of force balance for a single vortex takes the form
\be
\epsilon^{ijk} \hat{\kappa}_j (v^\N_k - v^\f_k) + \mathcal{F}^i = 0
\ee
As long as the system remains below the unpinning limit, this provides us with the required pinning force $\mathcal{F}^i$.
Given this, we find that the force acting on the neutrons is:
\be
{ f_\f^i \over \rho_\f \kappa}=-\epsilon^{ijk} \hat{\kappa}_j (v^\N_k - v^\f_k)
\ee
while the reaction force on the charged component will be
\be
{ f_\N^i \over \rho_\f \kappa}=\epsilon^{ijk} \hat{\kappa}_j (v^\N_k - v^\f_k)
\ee
The main conclusion from this exercise is that, if the lag between the two components is sufficiently small to allow us to consider the vortices as perfectly ``pinned'',
the exact form of the pinning force does not appear explicitly in the equations of motion.

\subsection{Vortex creep}

Let us now consider the situation where the lag between the two components is close to the critical value for unpinning. This regime is of great physical interest as it is at the heart
of many theoretical models for pulsar glitches. If one assumes that parts of the system are always slightly subcritical, one may
allow for a population of thermally  excited vortices to unpin randomly and transfer angular momentum to the crust. This is usually refered to as ``vortex creep'' \citep{alpar84}.
Describing this behaviour is clearly a challenging task, both from the microscopic and the macroscopic point of view.
On the one hand, there have been efforts to calculate the pinning ``force'' and the barrier that the thermally excited
vortices would have to overcome to unpin \citep{linkpaper}.
On the other hand, there have been attempts to incorporate the concept of vortex creep in a
macroscopic hydrodynamical description, by assuming that only a fraction of the vortices, on average, participates in the dynamics \citep{jahanmiri}.

Here we adopt a phenomenological approach aimed at exploring the hydrodynamics of the creep regime.
We start by noting that vortex creep would correspond to motion such that $v_\v^i \approx v_\N^i$. Assuming that \eqref{acomp} describes the ``pinning force''
completely, i.e. that there is no component along $\hat{\kappa}^i$, we can rewrite \eqref{vv} as
\begin{multline}
v_\mathrm{cr}^i = v_\v^i - v_\N^i
=  {1 \over 1 + \mathcal{R}^2 } \left[ a_\parallel \mathcal{R} - w_{\N\f} - a_\perp \right]  \hat{w}_{\N\f}^i
\\
+ {1 \over 1 + \mathcal{R}^2 } \left[ a_\parallel +  \left( w_{\N\f} + a_\perp \right) \mathcal{R}  \right]  \epsilon^{ijk} \hat{\kappa}_j \hat{w}^{\N\f}_k
\end{multline}
From this expression we see that there are two ways of enforcing vortex creep. One would be to let $\mathcal{R} \to \infty$. This  model has recently been
used in studies of precession and the unstable r-modes \citep{glitchlett}. However, with the ``pinning'' force explicitly in the problem  we have another option.
Focussing on the $\mathcal{R}\ll 1$ case, which should be relevant in the crust \citep{feibelman,bildsten}, we can demand that
\be
a_\parallel \mathcal{R} -  w_{\N\f} - a_\perp \equiv v_\parallel \ll v_\N
\ee
and
\be
 a_\parallel +  \left( w_{\N\f} + a_\perp  \right) \mathcal{R} \equiv v_\perp  \ll v_\N
\ee
With these definitions the creep velocity is given by
\be
v^i_\mathrm{cr} = v_\parallel \hat{w}^i + v_\perp \epsilon^{ijk} \hat{\kappa}_j \hat{w}_k
\ee
and we have
\be
v^2_\mathrm{cr} = v_\parallel^2 + v_\perp^2
\ee
This model is, obviously, more complicated since we now have three coefficients to specify; we need $\mathcal{R}$, $a_\parallel$ and $a_\perp$. However,
this provides more flexibility and
could allow us to, for example, consider a specific form for the ``pinning'' force or, indeed, the creep rate $v_\mathrm{cr}$. An important point is that, in this model you do not
have to have strong drag, $\mathcal{R} \gg 1$, to effect pinning. This may be particularly
relevant if we want to make our neutron star precession models more realistic,see \citet{link}.

Let us conclude by writing down the force that enters in the hydrodynamics.
After some straightforward algebra we have
\be
a_\parallel \approx v_\perp + v_\parallel \mathcal{R}
\ee
and
\be
a_\perp \approx -w - v_\parallel + v_\perp \mathcal{R}
\ee
This means that the force becomes
\be
{ f_\N^i \over \rho_\n \kappa}  \approx - v_\perp \hat{w}^i + (w + v_\parallel) \epsilon^{ijk} \hat{\kappa}_j \hat{w}_k
\ee
which means that
\be
f_\N^2 \approx v_\perp^2 + \left( w+v_\parallel\right)^2 \approx w^2
\ee
The required hydrodynamical force follows once we multiply by $\rho_\f n_\v \kappa$.

It is important to remember that the ``pinning'' force discussed here does not describe the realistic interaction between the vortices and the nuclei.
We have discussed a purely phenomenological model which  allows us to describe the motion when vortices are not yet completely free so that the drag
force cannot be approximated as linear in the velocities \citep{linkpaper}. Our simple ``pinning'' model
allows us to consider the dynamical implications of this regime.

\section{Lagrangian perturbations}

So far, we have discussed the general conditions that prevail in the inner neutron star crust and a two-fluid ``hydrodynamics''
model that accounts for the key features. We will now take an important step towards astrophysical applications by developing a
framework for Lagrangian perturbations of this system. The aim is to provide a model that can be applied to a range of important problems in
neutron star dynamics, from pulsar glitches to the gravitational-wave driven instability of the r-modes and magnetar oscillations.
These problems are all naturally approached within perturbation theory. Moreover, they all represent scenarios where the
sensitive interplay between the crust, the superfluid and the magnetic field is expected to be important.

\subsection{The unentrained two-fluid problem revisited}
\label{unentrained}

Since the two-fluid problem has two  dynamical degrees of freedom it is natural to introduce
two distinct Lagrangian displacement vectors $ {\xi}^i_\X $ \citep{kirsty}. In order to distinguish between these
displacements, we use
variations $\Delta_\X$ such that (for any, scalar or vectorial, quantity $Q$)
\beq
\Delta_\X Q = \delta Q + \mathcal{L}_{\xi_\X} Q \ ,
\eeq
where $\delta$ represents an Eulerian perturbation.
The perturbed continuity equations, c.f. \eqref{conti}, then take the form \citep{fs78}
\beq
\Delta_\X  n_\X = - n_\X \nabla_i \xi_\X^i \longrightarrow
\delta n_\X = - \nabla_i (n_\X \xi_\X^i) \ .
\label{super:continuity}
\eeq
This means that the equation that describes the perturbed gravitational potential is
  \be
  \nabla^2 \delta \Phi
  = 4 \pi m_\b G (\delta n_\X + \delta n_\Y)\\
  = - 4 \pi m_\b G \nabla_i (n_\X \xi^i_\X + n_\Y \xi^i_\Y)
\label{super:poisson}
  \ee

Considering the simplest case of vanishing entrainment and no ``external'' forces, $f_i^\f=f_i^\N=0$, the
results of \citet{kirsty} show that
\be
  \left( \partial_t   + \mathcal{L}_{v_\X}\right)\Delta_\X v^\X_i   +  \nabla_i
       \left(\Delta_\X \Phi + \Delta_\X  \tilde{\mu}_\X - { 1\over 2} \Delta_\X  v_\X^2\right)
 = 0
\label{sfpeul1}\ee
After some algebra, this leads to
\begin{multline}
\partial_t^2 \xi^\X_i +
2 v_\X^j \nabla_j \partial_t \xi^\X_i +  (v_\X^j \nabla_j)^2 \xi^\X_i +
\nabla_i \delta \Phi \\ +
\xi_\X^j \nabla_i \nabla_j \Phi  
- (\nabla_i \xi_\X^j) \nabla_j\tilde{\mu}_\X
+\nabla_i \Delta_\X \tilde{\mu}_\X
  = 0 \ .
\label{sfpeul2}
\end{multline}
Here, the Lagrangian perturbation of the chemical potential can be written  (with $\Y\neq\X$)
\begin{multline}
  \Delta_\X \tilde{\mu}_\X
  =  \delta \tilde{\mu}_\X
  +  \xi^j_\X \nabla_j \tilde{\mu}_\X \\
=  \left( {\partial \tilde{\mu}_\X \over \partial n_\X}
 \right)_{n_\Y} \delta n_\X
  + \left( {\partial \tilde{\mu}_\X \over \partial n_\Y}
 \right)_{n_\X} \delta n_\Y
  +  \xi^j_\X \nabla_j \tilde{\mu}_\X   \\
  =  -  \left( {\partial \tilde{\mu}_\X \over \partial n_\X}
 \right)_{n_\Y}
  \nabla_j (n_\X \xi^j_\X)
  - \left( {\partial \tilde{\mu}_\X \over \partial n_\Y}
 \right)_{n_\X}
  \nabla_j (n_\Y \xi^j_\Y) +  \xi^j_\X \nabla_j \tilde{\mu}_\X
  \end{multline}
 using the fact that $\tilde{\mu}_\X = \tilde{\mu}_\X (n_\f, n_\N)$.
Hence, we arrive at the following final form for the perturbed
Euler equations;
\begin{multline}
\partial_t^2 \xi^\X_i +
2  v_\X^j \nabla_j \partial_t \xi^\X_i + (v_\X^j \nabla_j)^2 \xi^\X_i +
 \nabla_i \delta \Phi
+ \xi_\X ^j \nabla_i \nabla_j (\Phi + \tilde{\mu}_\X)
 \\
- \nabla_i \left[ \left( {\partial \tilde{\mu}_\X \over \partial n_\X}
 \right)_{n_\Y} \nabla_j (n_\X\xi_\X^j) +
\left( {\partial \tilde{\mu}_\X \over \partial {n_\Y}} \right)_{n_\X}
\nabla_j (n_\Y\xi_\Y^j)
\right]  = 0
\label{sfpeul3}
\end{multline}

\citet{kirsty} demonstrated how one can proceed further and derive useful conserved quantities,
extending the single-fluid analysis of \citet{fs78} to the two-fluid arena. The importance of the
obtained canonical
energy $E_c$ stems from the fact that it can be used to
assess the stability of the system. In order for the evolution to be dynamically unstable, i.e.
for a perturbation to blow up in absence of additional forces, we must have $E_c=0$.
A secular (viscosity or radiation driven) instability requires $E_c<0$, provided that the energy lost
through dissipation is positive (which makes sense). A particularly nice feature of the
analysis of \citet{kirsty} was the proof that the standard instability criterion for gravitational-wave instabilities, that a normal mode of oscillation becomes
unstable when its pattern speed changes sign (eg. when an originally backwards retrograde mode in a rotating
star becomes prograde \citep{gwreview}), holds also in the two-fluid problem. This had previously been assumed to be the case,
but there was no formal proof.
In the following, we will not attempt to address the issue of the canonical energy
for more complex systems; our focus is entirely on the perturbed equations of motion. A stability
analysis for these equations would be interesting, but as this may well be prohibitively complicated we
leave the problem for future considerations.

\subsection{Accounting for elasticity and the magnetic field}

The equation that represents that crust dynamics must account for both elastic and magnetic contributions.
The former leads to the well-known contribution
\be
\Delta_\N \left( f^\N_i/\rho_\N \right) = { 1 \over \rho_\N} \nabla^j  \sigma_{ij}
\ee
where the shear tensor is given by
\be
\sigma_{ij} = \check \mu ( \nabla_i \xi^\N_j + \nabla_j \xi^\N_i) - { 2 \over 3} \check \mu (\nabla^l \xi_l^\N) \delta_{ij}
\ee
(here one should not confuse the shear modulus $\check \mu$  with the chemical potentials $\mu_\X$). It is important to keep in mind that
these expressions are only valid for unstrained background configurations.

In the case of the magnetic field, we need the Lagrangian perturbation of \eqref{lorentz1}.
The required results have already been derived by \citet{maglag}, so we simply restate them here.
The perturbations are determined from
\be
\Delta_\N \left( {B^j \over \rho_\N} \right) = 0
\label{ind1}
\ee
which leads to
\be
\Delta_\N B^i = -B^i \nabla_j \xi^j_\N
\label{db1}\ee
and
\be
\Delta_\N B_i = B_j \nabla_i \xi_\N^j - B_i (\nabla_j \xi_\N^j) + B^j \nabla_j \xi^\N_i
\label{db2}\ee
Finally, using
\be
\Delta_\N ( \nabla_j B_i) = \nabla_j ( \Delta_\N B_i) - B_l \nabla_j \nabla_i \xi_\N^l
\ee
we obtain from (\ref{lorentz1})
\be
\Delta_\N \left( f^\N_i/\rho_\N \right)  = {B^j \over 4 \pi \rho_\N} [ \nabla_j ( \Delta_\N B_i) - \nabla_i (\Delta_\N B_j)]
\ee
It is straightforward to express the magnetic perturbations in terms of the displacement $\xi_\N^i$, but since expressions that result are
quite involved we have decided not to do so here.

\subsection{Entrainment}

The elastic and magnetic forces take relatively simple forms. However, the fluid part of the problem becomes much more complex
if we consider the generic situation where the entrainment is not vanishing.
In the unentrained case the two equations of motion are
coupled \underline{chemically} through the equation of state (e.g. through various interactions) and \underline{gravitationally}
since variations in the number density
of one fluid affect the gravitational potential and hence the motion of the other fluid.
In contrast, the entrainment parameter $\alpha$
encodes how the internal energy of the system depends on the \underline{relative velocity} of the two fluids. This usually leads to a stronger coupling of the components.
Since the entrainment encodes the effective dynamical mass of each matter constituent, it will affect most scenarios that involve
the superfluid regions of the star.

Including the entrainment, the Euler equations take the form \eqref{Eulers}. However, since
\beq
v_\X^j \nabla_j \( \eps_\X \wid \)
= \mathcal{L}_{v_\X}\(\eps_\X \wid\) - \eps_\X \wjd \nabla_i v_\X^j
\eeq
these can be rewritten as
\beq
  \left(\partial_t + \mathcal{L}_{v_\X} \right) \left(v^\X_i + \eps_\X w_i^{\Y\X} \right)
+ \nabla_i \left(\Phi + \mut_\X  - \frac{v_\X^2}{2} \right) = 0
\label{euler2}\eeq
We want to  consider Lagrangian perturbations  of this system.
The derivation follows
the same approach as in the unentrained case, described by \citet{kirsty} and summarised in Section~\ref{unentrained}.

First we note that the continuity equations and the Poisson equation
are not affected by entrainment, so we can still use equations (\ref{super:continuity}) and (\ref{super:poisson}).
Secondly, perturbing the Euler equations we have
\begin{multline}
  \left(\partial_t + \mathcal{L}_{v_\X} \right) \left[ \Delta_\X v^\X_i + \Delta_\X \(\eps_\X w_i^{\Y\X} \) \right] \\
+ \nabla_i \left(\Delta_\X \Phi + \Delta_\X \mut_\X  - \frac{\Delta_\X v_\X^2 }{ 2} \right) = 0
\label{lageqn}
\end{multline}
This follows immediately since the Lagrangian variation commutes with $\partial_t + \mathcal{L}_{v_\X} $.
Most of the terms in this equation were already considered in the unentrained problem.
A key difference now is that $ \delta \tilde{\mu}_\X$ depends on the entrainment. This is obvious
since the chemical potential, $\mu_\X$, is the
partial derivative of the energy functional, $E$,  with respect to the  number density, $n_\X$. Since $E$ depends on the
entrainment the Eulerian variation of the chemical potential also depends on the entrainment. In general, we have
\begin{multline}
\delta \tilde{\mu}_\X = - \( {\partial \tilde{\mu}_\X \over \partial n_\X}\)_{n_\Y, w^2}\nabla_j \(n_\X \xi_\X^j \) \\
- \( {\partial \tilde{\mu}_\X \over \partial n_\Y}\)_{n_\X, w^2}\nabla_j \(n_\X \xi_\X^j \)
+ \( {\partial \tilde{\mu}_\X \over \partial w^2}\)_{n_\X,n_\Y} \delta w^2
\end{multline}
where
\beq
\( {\partial \tilde{\mu}_\X \over \partial w^2}\)_{n_\X,n_\Y}
= {1 \over m_\b } \( {\partial \alpha \over \partial n_\X}\)_{n_\Y, w^2} \equiv   {1 \over m_\b } \mathcal{A}_\X
\eeq
and
\beq
\delta w^2 = 2 \wjd \delta \wju
\eeq
giving
\begin{multline}
\delta \tilde{\mu}_\X = - \( {\partial \tilde{\mu}_\X \over \partial n_\X}\)_{n_\Y, w^2}\nabla_j \(n_\X \xi_\X^j \) \\
- \( {\partial \tilde{\mu}_\X \over \partial n_\Y}\)_{n_\X, w^2}\nabla_j \(n_\Y \xi_\Y^j \)
+ {2 \over m_\b} \mathcal{A}_\X \wjd \delta \wju
\label{eq:cf1}
\end{multline}
At this point it is worth noting that
\begin{multline}
\delta w_{\Y\X}^i =
\partial_t \xi_\Y^i + v_\Y^j \nabla_j \xi_\Y^i - \xi_\Y^j \nabla_j v_\Y^i
- \partial_t \xi_\X^i - v_\X^j \nabla_j \xi_\X^i + \xi_\X^j \nabla_j v_\X^i \\
= \partial_t \left( \xi_\Y^i - \xi_\X^i \right) + v_\X^j \nabla_j  \left( \xi_\Y^i - \xi_\X^i \right) \\
-  \left( \xi_\Y^j - \xi_\X^j \right) \nabla_j v_\X^i
+ w_{\Y\X}^j \nabla_j \xi_\Y^i - \xi_\Y^j \nabla_j w_{\Y\X}^i \\
= \left( \partial_t + \mathcal{L}_{v_\X} \right)  \left( \xi_\Y^i - \xi_\X^i \right) -  \mathcal{L}_{w_{\Y\X}}  \xi_\Y^i
\end{multline}

The only other piece of equation \eqref{lageqn} that was not present in the unentrained problem  can be written
\begin{multline}
\left(\partial_t + \mathcal{L}_{v_\X} \right)\Delta_\X \(\eps_\X \wid \) \\
= \eps_\X \left(\partial_t + \mathcal{L}_{v_\X} \right)\Delta_\X  \wid
+ \wid  \left(\partial_t + \mathcal{L}_{v_\X} \right)\Delta_\X  \eps_\X
\label{eq:remain}
\end{multline}
The first term follows easily from
\be
\Delta_\X  \wid = \delta w_i^{\Y\X} + \xi^j_\X \nabla_j w^{\Y\X}_i + w^{\Y\X}_j \nabla_i \xi_\X^j
\ee
The second term in \eqref{eq:remain} is worth discussing in more detail.
We first consider
\beq
\Delta_\X \eps_\X =
\delta \eps_\X  + \xi_\X^j \nabla_j \eps_\X
\eeq
Using the definition for $\eps_\X$, and the perturbed continuity equation, we find
\be
\Delta_\X \eps_\X =
{2 \over \rho_\X} \left[ \delta \alpha + \nabla_i \(\alpha \xi_\X^i \) \right]
\ee
In the general case, the entrainment parameter
$\alpha$ is a function of the two number densities, e.g. $n_\f$ and $n_\N$, and $w^2$. This means that
\beq
\delta \alpha = \mathcal{A}_\f \delta n_\f + \mathcal{A}_\N \delta n_\N + 2 \mathcal{A}_w w^{\N\f}_j \delta w_{\N\f}^j
\eeq
where we have defined
\beq
\label{eq:aw}
\mathcal{A}_w = \({\partial \alpha \over \partial w^2 } \)_{n_\f, n_\N}
\eeq
We can also use
\beq
  \nabla_i \alpha = \mathcal{A}_\f \nabla_i n_\f + \mathcal{A}_\N \nabla_i n_\N + 2 \mathcal{A}_w w_j^{\N\f} \nabla_i w_{\N\f}^j
\eeq
This means that
we can write $\delta \alpha$ as,
\begin{multline}
\delta \alpha = - \mathcal{A}_\f \nabla_i \(n_\f \xi_\f^i \) - \mathcal{A}_\N \nabla_i \( n_\N \xi_\N ^i \)
\\
+ 2 \mathcal{A}_w w^{\N\f}_j \[ \partial_t \xi_\N^j - \partial_t \xi_\f^j + v_\N^i \nabla_i \xi_\N^j
- v_\f^i \nabla_i \xi_\f^j - \xi_\N^i \nabla_i v_\N^j + \xi_\f^i \nabla_i v_\f^j \]
\end{multline}
After some algebra, we finally find that
\begin{multline}
\Delta_\X \eps_\X =
{2 \over \rho_\X} \Big\{ \(\alpha - \mathcal{A}_\X n_\X \) \nabla_i \xi_\X^i
- \mathcal{A}_\Y n_\Y \nabla_i \xi_\Y^i \\
+ 2 \mathcal{A}_w \wjd \left[ \partial_t \left( \xi_\Y^j - \xi_\X^j \right) - v_\X^i \nabla_i \xi_\X^j + v_\Y^i \nabla_i \xi_\Y^j
- \left( \xi_\Y^i - \xi_\X^i \right)\nabla_i v_\Y^j  \right] \Big\}
\label{dex}\end{multline}
We can now combine the above results to get a general expression for the right-hand side of \eqref{eq:remain}
in terms of the two displacements. However, this expression will be rather lengthy and may not be particularly useful.
In most situations of interest a reduced version should suffice. In principle, one may consider different
simplifying assumptions. The most drastic would be to consider the entrainment parameter to be uniform.
The natural way to achieve this would be to take $\alpha=$~constant\footnote{It is worth noting that $\eps_\X=\ $constant is only consistent for a
uniform density model.}. Then we have  $\mathcal{A}_\X = \mathcal{A}_w =0$, and
the equations simplify greatly. In fact, from \eqref{dex} we are only left with
\beq
\Delta_\X \eps_\X =
{2 \alpha \over \rho_\X} \nabla_i \xi_\X^i = \eps_\X  \nabla_i \xi_\X^i
\eeq

A more realistic model would be based on an expansion for small relative velocities \citep{gregexpand}. One would expect $w_{\Y\X}$ to be
small in most cases, so it makes sense to use the approximate equation of state
\beq
E(n_\f, n_\N, w^2) \approx E_0 (n_\f, n_\N) + E_1 (n_\f, n_\N) w^2
\eeq
In this case, we simply have
\beq
\alpha = E_1
\eeq
and it is obviously the case that $\mathcal{A}_w=0$. Hence, we have
\beq
\Delta_\X \eps_\X =
{2 \over \rho_\X} \left[ \(\alpha - \mathcal{A}_\X n_\X \) \nabla_i \xi_\X^i
- \mathcal{A}_\Y n_\Y \nabla_i \xi_\Y^i \right]
\eeq
This expression completes our analysis of the perturbations of the inviscid problem for the coupled
crust-superfluid system.

\subsection{Perturbing the mutual friction}

The various contributions associated with the superfluid vortices add further complexity to the
problem. As an illustration of this we will focus on the mutual friction,
which follows from equation \eqref{f1}. To complete the perturbation equations, we need 
\be
\Delta_\X\left(\frac{f^i_\N}{\rho_\f}\right)=-\Delta_\X \left(\frac{f^i_\f}{\rho_\f}\right)=
\Delta_\X \left(\mathcal{B}'_\mathrm{eff} \epsilon^{ijk} n_\v {\kappa}_j {w}_k - \kappa n_\v \mathcal{B}_\mathrm{eff} {w}^i\right)
\ee
where it is worth recalling that we defined
$\mathcal{W}_\f^i = n_\v \kappa^i$ in Section~2.2.

From equations (\ref{bb1}) and (\ref{bb2}) we see that the perturbations of the coefficients
$\mathcal{B}_\mathrm{eff}$ and $\mathcal{B}'_\mathrm{eff}$ will (in general) require a knowledge of
$\mathcal{R}$, $a_\parallel$ and $a_\perp$. The perturbations of these quantities can, obviously, be treated
in the same way as $\alpha$ in the previous section. However, in this case any simplifications would rely
on an understanding of the detailed microphysics. In addition to these quantities we need the variation
of the magnitude of the relative velocity, $w$. Determining this quantity is straightforward.

The main new piece of information required for the mutual friction is the perturbed vorticity.
In general, when the vortices are not moving with either of the macroscopic fluids, we need to perturb 
\eqref{circolo}, after solving for the vortex velocity as in Section~3. The procedure is relatively straightforward, 
but as the final expressions are  messy, and not very instructive, we will not work out the details here.  
Instead we consider the two extremes of  free and pinned vortices. 
 
In the first case, when the vortices are free so that  $v_\v^i = v_\f^i$, we see that \eqref{circolo} leads to
\beq
\left(\partial_t + \mathcal{L}_{v_\f} \right) \mathcal{W}_\f^i + \mathcal{W}_\f^i (\nabla_j v_\f^j ) = 0 \ .
\eeq
Perturbing this, it is quite easy to show that (in the case of a stationary and axisymmetric background)
\beq
\left(\partial_t + \mathcal{L}_{v_\f} \right) \left[ \Delta_\f \mathcal{W}_\f^i + \mathcal{W}_\f^i (\nabla_j \xi_\f^j ) \right] = 0
\eeq
We need the trivial solution to this equation, which means that we have
\beq
\Delta_\f \mathcal{W}_\f^i = -  \mathcal{W}_\f^i (\nabla_j \xi_\f^j )
\eeq

Finally, we perturb \eqref{circolo2} (which is completely analogous to the continuity equation for $n_\X$)
to get
\beq
\Delta_\f n_\v = - n_\v \nabla_j \xi_\f^j \ .
\eeq

Given these results, and the discussion in Section~3.1, it is easy to
work out what happens when the vortices are (perfectly) pinned. In that case, we have $v_\v^i = v_\N^i$ and
as a result we find that
\beq
\Delta_\N \mathcal{W}_\f^i = -  \mathcal{W}_\f^i (\nabla_j \xi_\N^j )
\eeq
and
\beq
\Delta_\N n_\v = - n_\v \nabla_j \xi_\N^j \ .
\eeq

\section{The crust-core interface}

In order for the developed perturbation framework to be useful for neutron star astrophysics, we need to
consider the crust-core interface. This region is known to be
important for a range of problems, especially since the associated viscous boundary layer may
provide efficient dissipation of large-scale flows in the core. We will not consider the viscous problem here,
but it is worth keeping in mind that a key issue concerns to what extent the velocity
perturbations are continuous across the interface. If they are not, then viscosity works to smooth out
the discontinuities (over some relatively short length scale) leading to damping of the bulk motion.
The magnetic field may play a similar role. When a magnetic field penetrates the interface, discontinuities
would induce Alfv\'en waves which would effect an efficient coupling between the crust and the core.

Another issue arising in this context is the potential appearance of nuclear pasta, that the nucleons form non-trivial topological clusters (e.g. rods or plates) rather than spherical nuclei arranged in a Coulomb lattice. Accounting for these structures is, in principle, straightforward once the properties of the various phases are understood. In most cases one would expect the system to remain ``isotropic'' on macroscopic scales owing to the fact that the pasta structures will ``freeze'' in a random fashion on some smaller scale, and the ``fluid model'' arises from a larger scale average. However, there are cases where these structures may be aligned on macroscopic scales (for instance due to a strong magnetic field or the existence of an ordered array of vortices). Then we may need to consider non-isotropic elasticity. The computation of the microscopic input parameters (equation of state, entrainment, shear modulus, et c.) is very complicated in the pasta phase. It may, for instance, be that the densities exhibit discontinuous jumps across the crust-core interface, which could lead to discontinuities in the velocities [see e.g.\ equation \eqref{con1} below] and thus to enhanced viscous damping as discussed above. These are very important issues, but for simplicity we will ignore them in the following, assuming an isotropic solid and continuous densities across the crust-core interface.

\subsection{Chemical gauge}

As before, we consider a system where a charge-carrying component is coupled to a neutral superfluid.
Furthermore, we assume that the superfluid extends across the interface.
At this point we have to return to the issue of the ``chemical gauge''. That is,
we have to discuss the physical meaning of $n_\f$ and $n_\N$. In developing the model, we have taken
the view that $n_\f$ represents the neutrons that are not confined to nuclei in the crust, while
$n_\N$ represents all protons as well as the confined neutrons (making up the ions in the lattice).
This view leads to a natural description of the elastic and magnetic forces. However, it does not
lead to a straightforward connection to the core, where one would usually distinguish \underline{all} the
neutrons, $n_\n$, from the protons, $n_\p$. Problems arise from the fact that the analysis
 requires variables that are ``meaningful'' across the crust-core interface. In principle, the problem would be
 more straightforward if we were to use a two-fluid model based on $n_\n$ and $n_\p$ also in the crust.
The downside to this would be that we would then have to reconsider the Lorentz force and the elasticity contributions.
After all, some of the neutrons will be associated with the  nuclei and hence should be affected by the crust motion.

Focussing on the generic case, we will connect the standard two-fluid model for the core with the
crust model we have developed. This forces us to consider  the relevance  of the chemical gauge
and serves to clarify some of the key issues.

The chemical gauge choice relates to the neutrons that are considered ``free''.
The issue is subtle since, in a dynamic situation, even the neutrons that are associated with the nuclei may be able to tunnel through
the relevant interaction potential. This makes concepts like the atomic number somewhat hazy. In general, one may introduce a new
basis such that
\be
n_\f^i = n_\n^i + (1-a_\N) n_\p^i
\ee
where $a_\N$ (which we will take to be constant in the following, a good approximation at the level of the individual fluid elements)
accounts for the fact that some of the neutrons move with the
(crust) protons. We also have
\be
n_\N^i = a_\N n_\p^i
\ee
Given these relations, it is easy to show that the neutron momentum is independent of the chemical gauge \citep{cacha}.
This follows immediately from the definition of the momentum \citep{monster};
\be
p^\X_i = {\partial \mathcal{L} \over \partial n_\X^i}
\ee
where $\mathcal{L}$ represents the relevant Lagrangian.
Hence, we have
\be
p^\n_i = p^\f_i
\ee
It also follows that
\be
\mu_\n = \mu_\f
\ee
However, these results also show that in general we must have $v_\f^i \neq v_\n^i$ and $\varepsilon_\f \neq \varepsilon_\n$.
Finally, in order to consider the vortices across the interface, it is natural to assume that
\be
\mathcal{W}_\f^i = \mathcal{W}_\n^i
\ee
The behaviour of the vortices may, of course, be more complicated than this but it makes sense to first consider the simplest
``reasonable'' model.

\subsection{The background configuration}

Following \citet{maglag} we represent the moving interface by a level set of a scalar function $f$ which can be extended
in a smooth fashion. Expecting the interface to move with the charged component, we require
\be
[\partial_t + {\cal L}_{v_\N} ] f = 0  \ ,
\ee
from which it is easy to show that
\be
[\partial_t + {\cal L}_{v_\N}] \nabla_i f = 0   \ .
\ee
In other words, the gradient $\nabla_i f$ is constant in the frame moving with $v_\N^i$. From this, it follows that
the perturbation $\Delta_\N f$ satisfies
\be
[\partial_t + {\cal L}_{v_\N} ] \Delta_\N f = 0  \ .
\ee
The trivial solution to this equation is $\Delta_\N f=0$, which essentially means that a fluid element
at the original surface remains at the perturbed surface. The normal to the surface can obviously be taken to be
$N_i = \nabla_i f$, and hence we have the unit normal
\be
\hat{N}_i = { \nabla_i f / N } \ , \mbox{ where } \quad N = |\nabla f| = (g^{ij} \nabla_i f \nabla_j f)^{1/2}
\ee
This means that
\be
[\partial_t + {\cal L}_{v_\N} ] \hat{N}_i = N_i [\partial_t + {\cal L}_{v_\N} ] N \ .
\ee
which shows that, even though $\hat{N}_i$ is not preserved by the flow,
any change in the unit normal is parallel to the normal itself. We will use this fact later.

These considerations are quite general. However, in the problem of interest we may restrict ourselves to
configurations (at the unperturbed level) that are stationary and axisymmetric. These assumptions mean that
we have $\partial_t N_i=0$ and $N_i v_\X^i=0$. The latter represents a no-penetration condition, simply
stating that (in the background configuration) the core fluids do not migrate into the crust.

In order to obtain the interface conditions, we identify a small cylinder of fluid aligned
with the normal to the interface, $\hat{N}_i$. Integrating the various equations over this small volume,
we will be able to deduce the relevant conditions to impose. Carrying out this exercise we need to make sure that the
equations we consider are valid in both the crust and the core. Given this, it is
natural to take as our starting point the conservation of baryon number. Assuming that the problem is stationary, the core equation
\be
\nabla_i \left( n_\n v_\n^i + n_\p v_\p^i \right) = 0
\ee
matches to the crust result;
\be
\nabla_i \left( n_\f v_\f^i + n_\N v_\N^i \right) = 0
\ee
Integrating these over the small volume, we see that we should impose
\be
\hat{N}_i \left( n_\n v_\n^i + n_\p v_\p^i \right) = \hat{N}_i \left( n_\f v_\f^i + n_\N v_\N^i \right)
\ee
at the interface. Noting that there are no chemical gauge issues concerning the protons, the
corresponding conservation law leads to (assuming that the densities are continuous
across the interface);
\be
\hat{N}_i \left( v_\p^i - v_\N^i \right) = 0
\label{con1}\ee
Given this, and the fact that the total number density is given by $n=n_\p+ n_\n= n_\f + n_\N$,
we can rewrite the first condition as
\be
n_\n \hat{N}_i \left( v_\n^i - v_\p^i \right) = n_\f \hat{N}_i \left( v_\f^i - v_\N^i \right)
\label{con2}\ee
The final conditions \eqref{con1} and \eqref{con2} are, of course, trivially satisfied for an axisymmetric system since $\hat{N}_i v_\X^i = 0$.

Moving on to the momentum equations, it is natural to work in the frame moving with the protons/crust. After all, the volume that we integrate
over is fixed in this frame. We can also safely treat any relative flow as planar, since the volume we consider is arbitrarily small.
These assumptions simplify the analysis greatly.

Let us first introduce the total momentum flux;
\be
\pi_i = \rho_\X v^\X_i + \rho_\Y v^\Y_i
\ee
where $(\X,\Y)$ is either $(\n,\p)$ or $(\f,\N)$, depending of whether we consider the core or the crust.
Combining the Euler equations in the appropriate way we find that \citep{monster}
\be
\partial_t \pi_i + \nabla_j \left( v_\X^j \pi^\X_i + v_\Y^j \pi^\Y_i \right) + \nabla_i p + \rho\nabla_i\Phi  = \nabla^j T_{ij} \ ,
\label{momtot}\ee
where the pressure $p$ is defined such that 
\be
\nabla_i p = \sum_\X n_\X\nabla_i \mu_\X  - \alpha \nabla_i w_{\X\Y}^2 \ .
\ee
The elastic and magnetic stresses are accounted for in $T_{ij}$.

Let us consider the problem on the crust side of the interface (the core follows simply from letting $\f\to\n$ and $\N\to \p$).
In the frame moving with the crust we have $v_\N^i=0$ (obviously), and we also need to replace $v_\f^i\to w_{\f\N}^i$. This leads to
\be
\pi_i^\f \to \rho_\f (1-\varepsilon_\f) w^{\f\N}_i \ ,
\ee
and the momentum equation takes the form
\be
\partial_t \pi_i + \nabla_j \left[ \rho_\f (1-\varepsilon_\f) w_{\f\N}^j w^{\f\N}_i + \delta_i^j p \right] +\rho\nabla_i\Phi = \nabla^j T_{ij} \ ,
\ee
We now integrate this equation over the small volume straddling the interface. As long as the total density is continuous across the interface, the
gravitational potential and its derivative will be smooth and therefore the corresponding integral vanishes as we let the volume shrink.
This exercise tells us that there will be no local force associated with the interface as long as
\begin{multline}
\hat{N}_j \left[ \rho_\f (1-\varepsilon_\f) w_{\f\N}^j w^{\f\N}_i + \delta_i^j p - T_i^j \right]_\mathrm{crust}  \\
=
\hat{N}_j \left[ \rho_\n (1-\varepsilon_\n) w_{\n\p}^j w^{\n\p}_i + \delta_i^j p - T_i^j \right]_\mathrm{core}
\label{gencond}\end{multline}
This condition is quite general. In particular, it needs to hold also on the perturbative level. As far as the background
configuration is concerned, we obviously have $\hat{N}_j v_\X^j=0$, which means that we have (representing the change in a given quantity across the interface by
 $\langle \ldots \rangle$ \citep{maglag});
\be
\hat{N}_j \langle \delta_i^j p -   T_i^{\ j} \rangle  = 0  \ ,
\label{cond0}\ee
These are the usual traction conditions.

For the vertical component see that
\be
\langle p \rangle =  \hat{N}^i \hat{N}^j \langle T_{ij} \rangle \ ,
\label{cond1}\ee
while the horizontal components lead to
\be
 \perp^{li} \hat{N}^j \langle T_{ij} \rangle = 0 \ ,
\label{cond2}\ee
where  we have defined the projection (orthogonal to the normal)
\be
\perp^{li} =
 g^{li} - \hat{N}^l \hat{N}^i \ .
\ee
It is relevant to note that
the pressure may now be affected by the presence of a relative flow.

If we assume that the background configuration is such that the crust is relaxed,  we
only need to account for the magnetic stresses. Then we have
\be
T_{ij} = - g_{ij} {B^2 \over 8\pi} + { 1 \over 4 \pi} B_i B_j \ .
\ee
We also know that $\nabla_i B^i=0$, which implies that we must have
\be
\langle \hat{N}_i B^i \rangle = 0 \ .
\label{bone}\ee
In this case, condition \eqref{cond1} leads to
\beq
\langle p + {B^2 \over 8 \pi } \rangle = 0 \ ,
\eeq
while \eqref{cond2} becomes
\be
\left( \hat{N}_j B^j \right) \perp^{li} \langle B_i \rangle = 0 \ .
\ee
Combined with \eqref{bone}, this shows that if the magnetic field penetrates the interface then all
components of the background field $B_i$ must be continuous.

To complete the analysis of the interface, we need one more condition. We obtain this condition from the
momentum equation for the neutrons. This choice is natural since we need to consider a quantity that remains relevant on both sides of
the interface, and there are no chemical gauge issues concerning the momentum of the superfluid
 (essentially since it follows from the phase of the macroscopic
quantum wavefunction). However, we still have to be careful. Basically, the presence of vortices and potential pinning
complicates the picture. To make progress we will take the view that the irrotational condensate and the vortices can be considered separately.
At the end of the day, the total momentum equation involves an average over these components. In effect,  the interface
must reflect this large scale average. Our approach to the problem represents this, yet it is
admittedly rather naive. A number of issues need to be better understood, in particular concerning the way in which
vortices extend from the fluid core to the elastic environment of the crust.
Let us simply mention two problems: First of all, we know from low-temperature laboratory
superfluids that vortices connect orthogonally to solid walls. It is natural to ask if the
same is true for vortices that penetrate the neutron star crust-core interface. Secondly, the entrainment
in the core leads to the vortices being magnetised, due to the entrainment, but this effect
relies on the protons being superconducting so is not active in the crust. The upshot is that a magnetised v
ortex somehow connects to an unmagnetized one. How does this work? The answer may be linked to the
transition from superconducting protons to ones locked in nuclei. In this case, one would expect the
presence of a current sheet. Presumably, this may also resolve any issues concerning the magnetic vortices,
but the details have not yet been considered. These and other issues need to be resolved by
future work. In the following we will adopt the pragmatic view that really difficult problems are perhaps best ignored.

Anyway, considering first the irrotational part, we have
\be
\partial_t \tilde{p}^\X_i + \nabla_i \left( \tilde{\mu}_\X - {1\over 2} v_\X^2 + v_\X^j \tilde{p}^\X_j \right) = 0
\ee
where $\X=[\n,\f]$ depending on whether we are in the core or the crust. Working in the crust frame, and integrating over a small
volume, this leads to the interface condition
\be
\tilde{\mu}_\f - \left( {1 \over 2}-\varepsilon_\f\right) w_{\f\N}^2 = \tilde{\mu}_\n - \left( {1 \over 2}-\varepsilon_\n\right) w_{\n\p}^2
\label{mucon}\ee
It should be noted that, in the particular choice of chemical gauge where we consider all the neutrons in the crust
we have $\f\to\n$. In that case, the neutron chemical potential must be continuous, as expected.

Moving on to the vortices, we still do not have to worry about chemical gauge issues. Basically, one would expect each vortex to penetrate into the crust
leading to the vortex density $n_\v$ being continuous. Hence, we consider \eqref{circolo1},
where we recall that the vortex velocity $v_\v^i$ depends on whether there is pinning or not.
Working in the crust frame (as before) and integrating, we arrive at
the condition
\be
(\hat{N}_j \mathcal{W}_\f^j ) w_i^{\v\N} = (\hat{N}_j \mathcal{W}_\n^j ) w_i^{\n\p}
\ee
where we have used the fact that (in the present analysis) we do not consider pinning in the core.
However, because of the chemical gauge invariance, we have $\mathcal{W}_\f^j = \mathcal{W}_\n^j$ which means that
our final interface condition is
\be
 v_i^\v =  v_i^\n
\label{vortcon}\ee
This result is extremely intuitive. If the crust vortices are free, moving with $v_\f^i$, then we must have
$v_\f^i = v_\n^i$ at the interface. Meanwhile, if the vortices are pinned,  moving with $v_\N^i$, then the condition should be
$v_\N^i = v_\n^i$.

These results represent two limiting cases. In the general case, we would also need to keep track of the 
vortex velocity. This is, in principle, straightforward but we will not discuss the results here. It may be worth noting 
that such models may be considered from a three-fluid point-of-view, with the vortices forming a distinct ``species''.
This is an interesting strategy that could prove advantageous in some situations.

\subsection{The perturbed problem}

Let us now move on to the conditions that need to be imposed at the linear perturbation level.
We can think of two possible strategies. Either we take the view that the conditions derived in the previous
section are ``exact'', which means that we can perturb them directly, or we start from the relevant perturbation equations
and carry out the analysis for a small volume straddling the interface all over again. In principle, these two approaches should lead to
the same answer. In practice, we find it useful to use a combination of them.

We begin by considering particle conservation.
In general, Lagrangian variation of the continuity equations leads to
\be
\left( \partial_t + \mathcal{L}_{v_\X} \right) \left( \Delta_\X n_\X +  n_\X \nabla_i \xi_\X^i \right) = 0
\ee
We require the trivial solution, i.e. take
\be
\Delta_\X n_\X +  n_\X \nabla_i \xi_\X^i  = 0
\ee
In the case of the protons, we can integrate this equation over the small cylinder across the interface.
Taking the volume small enough that variations in $n_\p$ can be neglected (we are not allowing for
density discontinuities), we then find that
\be
N_i \left( \xi_\p^i - \xi_\N^i \right) = 0
\label{pcon10}\ee
As one would have expected, the normal component of the proton displacement should be continuous.
This condition follows immediately if we want to avoid there being a void (or overlap) in the proton fluid at the interface.

To get the second condition , we perturb the equation for total baryon conservation. This leads to (in the crust)
\be
\Delta_\N n = \Delta_\N (n_\f + n_\N) = - n_\N \nabla_j \xi_\N^j + \xi_\N^j \nabla_j n_\f - \nabla_j \left( n_\f \xi_\f^j \right)
\ee
If we (again) assume that the densities are smooth, then we only need to consider
\be
\Delta_\N n \approx - \nabla_j \left( n_\N \xi_\N^j +  n_\f \xi_\f^j \right)
\ee
After integration across the interface, we find that we must have
\be
N_j \left[ n_\n \left( \xi_\N^j - \xi_\n^j \right) - n_\f \left( \xi_\N^j - \xi_\f^j \right) \right]  = 0
\label{mixcon}\ee
In the case of the comprehensive gauge, when $n_\f=n_\n$, this reduces to
\be
N_j \left( \xi_\n^j - \xi_\f^j \right)  = 0
\ee
Again, this condition seems natural, and we learn that the more complicated nature of \eqref{mixcon} results from the
fact that not all neutrons in the crust are free. We also see that we must keep careful track of the
different number densities across the interface.

Next we need the perturbed versions of \eqref{cond1} and \eqref{cond2}.
To derive these, we assume (as in the standard level set method) that the general conditions can meaningfully be extended away from the interface, and be perturbed in the usual way. There may be some technical issues
associated with this approach, but we will not go into the details of this here.
To derive the relevant conditions we start from \eqref{gencond}
which leads to
\begin{multline}
 \hat{N}_j \left[ \rho_\f (1-\varepsilon_\f) w_i^{\f\N} \partial_t (\xi_\f^j - \xi_\N^j) + \delta_i^j \Delta_\N p -\Delta_\N T_i^{\ j} \right]  \\
 =
  \hat{N}_j \left[ \rho_\n (1-\varepsilon_\n) w_i^{\n\p} \partial_t (\xi_\n^j - \xi_\p^j) + \delta_i^j \Delta_\p p - \Delta_\p T_i^{\ j} \right] \ .
\end{multline}
It is easy to see that, for an axisymmetric background, the first term in each expression does not contribute to the
normal component. Contracting with $\hat{N}^i$ we are left with
\be
\left[ \Delta_\N p - \hat{N}^i \hat{N}_j \Delta_\N T_i^{\ j} \right]_\mathrm{crust}  =
\left[ \Delta_\p p - \hat{N}^i \hat{N}_j \Delta_\p T_i^{\ j} \right]_\mathrm{core} \ .
\ee
That is, we arrive at the expected traction condition.
The horizontal result is (obviously) more complicated. A projection orthogonal to $\hat{N}^i$ leads to
\begin{multline}
\hat{N}_j \left[ \rho_\f (1-\varepsilon_\f) w_i^{\f\N} \partial_t (\xi_\f^j - \xi_\N^j) - (\delta^l_i-\hat{N}^l \hat{N}_i)\Delta_\N T_l^{\ j} \right]_\mathrm{crust}  \\
=
  \hat{N}_j \left[ \rho_\n (1-\varepsilon_\n) w_i^{\n\p} \partial_t (\xi_\n^j - \xi_\p^j)  - (\delta^l_i-\hat{N}^l \hat{N}_i)\Delta_\p T_l^{\ j}  \right]_\mathrm{core} \ .
\end{multline}
As in previous cases, the additional complications arise from the choice of chemical gauge.

In the magnetic field case, we need \eqref{db1} and \eqref{db2}.
It is also useful to note that \eqref{bone} leads to
\be
\langle \hat{N}_i \Delta_\N B^i \rangle = 0 \ .
\ee
From this we see that
\be
\langle \hat{N}^i \Delta_\N B_i \rangle = \langle B^i \hat{N}^j \left( \nabla_i \xi^\N_j + \nabla_j \xi^\N_l \right) \rangle \ .
\ee
We then have
\begin{multline}
\hat{N}_i \Delta_\N p - \hat{N}_j \Delta_\N T_i^{\ j}\\
=  \hat{N}_i \left( \Delta_\N p + { 1 \over 8 \pi} \Delta_c B^2 \right) -
{1 \over 4 \pi} \hat{N}_j \left( B^j \Delta_\N B_i + B_i \Delta_\N B^j \right) \ .
\end{multline}
The vertical component becomes
\be
\langle \Delta_\N p + { 1 \over 8 \pi} \Delta_c B^2 \rangle = { 1 \over 4 \pi} \left( \hat{N}_j B^j \right) \langle \hat{N}^i \Delta_\N B_i \rangle  \ ,
\ee
while the horizontal condition becomes
\be
\perp^{li}\langle \hat{N}_j \left( B^j \Delta_\N B_i + B_i \Delta_\N B^j \right) \rangle = 0 \ .
\ee

At the perturbation level, we also need to consider the elastic problem. As before, we assume that the background
configuration is relaxed (and the core is fluid!), in which case we have 
\be
\Delta_\N T_{ij} = \check \mu \left( \nabla_i \xi^\N_j + \nabla_j \xi^\N_i \right) - {2 \over 3} \check \mu g_{ij} \nabla_l \xi_\N^l \ ,
\ee
and
\be
\Delta_\N T_i^{\ j} = g^{jl} \Delta_\N T_{il} \ .
\ee
In this case, the vertical condition becomes
\be
\langle \Delta_\N p + {2 \over 3} \check \mu \nabla_j \xi_\N^j \rangle = 2 \langle \check \mu \hat{N}^i \hat{N}^j \nabla_i \xi^\N_j \rangle \ ,
\ee
while the horizontal one can be written
\be
\perp^{li} \langle \check \mu \hat{N}^j \left( \nabla_i \xi^\N_j + \nabla_j \xi^\N_i \right) \rangle = 0  \ .
\ee

Combining the elastic and magnetic results to arrive at the conditions to impose in the general case is, of course, straightforward.

Finally, we consider the conditions relating to the superfluid component. Perturbing the scalar condition
\eqref{mucon} we see that we should require
\begin{multline}
\left[ \Delta_\N  \tilde{\mu}_\f  -  w_{\f\N}^2  \Delta_\N \varepsilon_\f + \left( {1 \over 2} - \varepsilon_\f \right) \Delta_\N w_{\f\N}^2 \right]_\mathrm{crust} = \\
\left[ \Delta_\p \tilde{\mu}_\n  -  w_{\n\p}^2  \Delta_\p \varepsilon_\n + \left( {1 \over 2} - \varepsilon_\n \right) \Delta_\p w_{\n\p}^2 \right]_\mathrm{core}
\end{multline}
As far as the vorticity is concerned, it follows naturally from the discussion leading up to \eqref{vortcon} that we should have
\be
\perp^i_j \left( \Delta_\N v_\v^j - \Delta_\p v_\n^j \right) = 0
\ee
This condition is (obviously) satisfied if we have $\perp^i_j(\xi_\f^j-\xi_\n^j)=0$ in the free vortex case and   $\perp^i_j(\xi_\N^j-\xi_\n^j)=0$ when perfect pinning prevails.
This ensures that there are no kinks in the vortices at the interface.

\section{Conclusions}

We have developed a Lagrangian perturbation framework relevant for the conditions 
that apply in a mature neutron star, accounting for the presence of superfluid components, 
the elastic crust and the magnetic field. The considered physics impacts on a wide range of astrophysical
phenomena, from pulsar glitches to magnetar seismology and various gravitational-wave emission mechanisms. 
Hence, the reported theoretical developments provide us with a solid foundation to consider exciting applications. 

There is also significant scope for future improvements of the theory. Most importantly, the effort should be extended to 
relativistic gravity. This would open the door to truly quantitative considerations of realistic neutron star models, e.g. 
based on a modern supranuclear equation of state with composition and thermal gradients. Developments in this 
direction are in progress. 

We also need to improve our understanding of the  physics involved. 
Our discussion highlights the need to know a wider set of parameters, like the superfluid entrainment both in the 
star's core and in the elastic crust. The various interactions involving superfluid vortices, from pinning to mutual friction, 
also need to be understood. While we can continue to advance our understanding of the phenomenology of these very complex systems, 
we need to impose realistic constraints on our models. This requires, if not a precise knowledge of the involved parameters, 
some idea of what the permissible ranges may be. To achieve this goal we need a continued dialogue across different branches of physics, 
a challenging but ultimately rewarding exercise.

\section*{Acknowledgments}

NA would like to thank Bennett Link for useful conversations on neutron star superfluids in general 
and vortex interactions in particular.
Our model for the pinning in the crust is heavily influenced by a talk that Bennett gave at the neutron star meeting at Nordita in Stockholm in 2009. 
We are also grateful to Pierre Pizzochero for sharing his thoughts on the nuclear pinning force.
NA acknowledges support from STFC via grant number PP/E001025/1. LS is supported by the European Research Council under Contract No.\ 204059-QPQV,
and the Swedish Research Council under Contract No.\ 2007-4422. BH is supported by a Marie Curie Fellowship.

\end{document}